\documentclass[conference]{IEEEtran}
\IEEEoverridecommandlockouts
\usepackage{cite}
\usepackage{gensymb}
\usepackage{amsmath,amssymb,amsfonts}
\usepackage{algorithmic}
\usepackage{graphicx}
\usepackage{textcomp}
\usepackage{xcolor}
\def\BibTeX{{\rm B\kern-.05em{\sc i\kern-.025em b}\kern-.08em
    T\kern-.1667em\lower.7ex\hbox{E}\kern-.125emX}}
\begin{document}

\title{Thermal Analysis of a 3D Stacked High-Performance Commercial
Microprocessor using Face-to-Face Wafer Bonding Technology\\
{}
}

\author{\IEEEauthorblockN{Rahul Mathur\IEEEauthorrefmark{1}\IEEEauthorrefmark{3},
Chien-Ju Chao\IEEEauthorrefmark{1},
Rossana Liu\IEEEauthorrefmark{1},
Nikhil Tadepalli\IEEEauthorrefmark{2},
Pranavi Chandupatla\IEEEauthorrefmark{1},
Shawn Hung\IEEEauthorrefmark{1},\\
Xiaoqing Xu\IEEEauthorrefmark{1},
Saurabh Sinha\IEEEauthorrefmark{1}, and
Jaydeep Kulkarni\IEEEauthorrefmark{3}}
\IEEEauthorblockA{\IEEEauthorrefmark{1}Arm Inc.,
5707 Southwest Parkway, Austin, TX, 78735\\
Email: rahul.mathur@arm.com}
\IEEEauthorblockA{\IEEEauthorrefmark{2}Arm Ltd, Cambridge, UK}
\IEEEauthorblockA{\IEEEauthorrefmark{3}University of Texas at Austin, TX, USA}}
\maketitle

\begin{abstract}

3D integration technologies are seeing widespread adoption in the semiconductor industry to offset the limitations and slowdown of two-dimensional scaling. High-density 3D integration techniques such as face-to-face wafer bonding with sub-10 $\mu$m pitch can enable new ways of designing SoCs using all 3 dimensions, like folding a microprocessor design across multiple 3D tiers. However, overlapping thermal hotspots can be a challenge in such 3D stacked designs due to a general increase in power density. In this work, we perform a thorough thermal simulation study on sign-off quality physical design implementation of a state-of-the-art, high-performance, out-of-order microprocessor on a 7nm process technology. The physical design of the microprocessor is partitioned and implemented in a 2-tier, 3D stacked configuration with logic blocks and memory instances in separate tiers (logic-over-memory 3D). The thermal simulation model was calibrated to temperature measurement data from a high-performance, CPU-based 2D SoC chip fabricated on the same 7nm process technology. Thermal profiles of different 3D configurations under various workload conditions are simulated and compared. We find that stacking microprocessor designs in 3D without considering thermal implications can result in maximum die temperature up to 12$\degree$C higher than their 2D counterparts under the worst-case power-indicative workload. This increase in temperature would reduce the amount of time for which a power-intensive workload can be run before throttling is required. However, logic-over-memory partitioned 3D CPU implementation can mitigate this temperature increase by half, which makes the temperature of the 3D design only 6$\degree$C higher than the 2D baseline. We conclude that using thermal-aware design partitioning and improved cooling techniques can overcome the thermal challenges associated with 3D stacking.
	\vspace{1pc}
	%\hline
	\vspace{1pc}

This is an accepted version of the IEEE published article presented at IEEE Electronic Components and Technology Conference (ECTC) 2020 http://www.ectc.net.

© 2020 IEEE. Personal use of this material is permitted. Permission from IEEE must be obtained for all other uses, in any current or future media, including reprinting/republishing this material for advertising or promotional purposes, creating new collective works, for resale or redistribution to servers or lists, or reuse of any copyrighted component of this work in other works.
\end{abstract}


\section{Introduction}
With the slowdown of Moore's Law of two-dimensional scaling \cite{moore:1965}, the semiconductor industry is at a critical juncture as 2.5D and 3D stacking technologies are being actively explored and adopted by certain end-applications \cite{yeric:2015}. 3D stacking enables vertical integration of more than one layer of active transistors and interconnects with the goal of increasing compute density. Integrated circuit (IC) designs with natural redundancy and regularity in 2D can be extended or stacked in the third dimension with relative ease. CMOS image sensors \cite{fontaine:2019}, DRAM memories \cite{jun:2017}, and NAND Flash memories \cite{venkatesan:2018} are all examples of ICs with high amounts of regularity and redundancy, and these products have already embraced 3D integration and achieved success in high-volume manufacturing.

However, the adoption of 3D stacking for logic applications has been relatively slow. Functionally complete chips commonly referred to as chiplets, are stacked using packaging technologies. The stacking configuration could be 2.5D, wherein, chiplets are assembled in 2D but interconnected through an underlying substrate, e.g., Silicon interposer or redistribution layer (RDL). Alternatively, the stacking configuration could be 3D, e.g., package-on-package (PoP) wherein DRAM packaged dies are stacked on ASIC die \cite{tseng:2016} or, two or more compute dies are stacked using through-silicon-via (TSV) and micro-bump technology \cite{Gomes:2020}.

Technology trends point towards finer-pitch 3D connectivity in the form of wafer or die-stacking 3D integration, which, we refer to as high-density 3D integration. Hybrid-bonding technologies allow precision alignment of wafers resulting in 3D connection pitches in the range of 10 $\mu$m \cite{chen:2019} to 1 $\mu$m \cite{Jouve:2017}. This technology opens up the possibility of designing systems where functional units are partitioned and co-designed across separate 3D tiers. The advantages of such 3D integration are multi-fold:

\vspace{-0.5mm}
\begin{itemize}
	\setlength\itemsep{0em}
\item Functional blocks separated across large distances in 2D can be brought closer in 3D, reducing interconnect delay and power.
\item Large-die SoCs can be partitioned into smaller dies, improving yield and potentially leading to lower cost.
\item Dies from different process technologies can be integrated enabling heterogeneous integration, leading to flexible product migration to advanced nodes for further cost reduction\cite{England:2017}.
\end{itemize}
\vspace{-0.5mm}

However, 3D stacking of compute dies can result in higher power density, leading to power delivery and thermal challenges. Performance throttling via dynamic voltage and frequency scaling (DVFS) can combat increasing on-die temperatures, but that would also offset some of the promised performance gains of 3D stacking. In this paper, we provide a comprehensive overview of the thermal impact of high-density 3D stacking technologies in the context of a design-aware partitioned and timing-optimized 3D high-performance microprocessor design. The main contributions of this paper are as follows.
%\vspace{-1.5mm}
\begin{itemize}
	\setlength\itemsep{0em}
\item To the best of our knowledge, this paper presents the first comprehensive thermal simulation study on a sign-off quality physical design of a 3D high-performance microprocessor using face-to-face (F2F) wafer-bonding technology.
\item The thermal simulation model is calibrated to 2D hardware measurement data from a high-performance CPU-based SoC. 
\item By evaluating various 3D stacking configurations in our study, we demonstrate that taking thermal implications into account early in the physical design process of a 3D partitioned microprocessor can partially mitigate the thermal impact of 3D stacking.
\end{itemize}
%\vspace{-1.5mm}

\begin{figure}[!t]
\centering
\centerline{\includegraphics[width=3.45in]{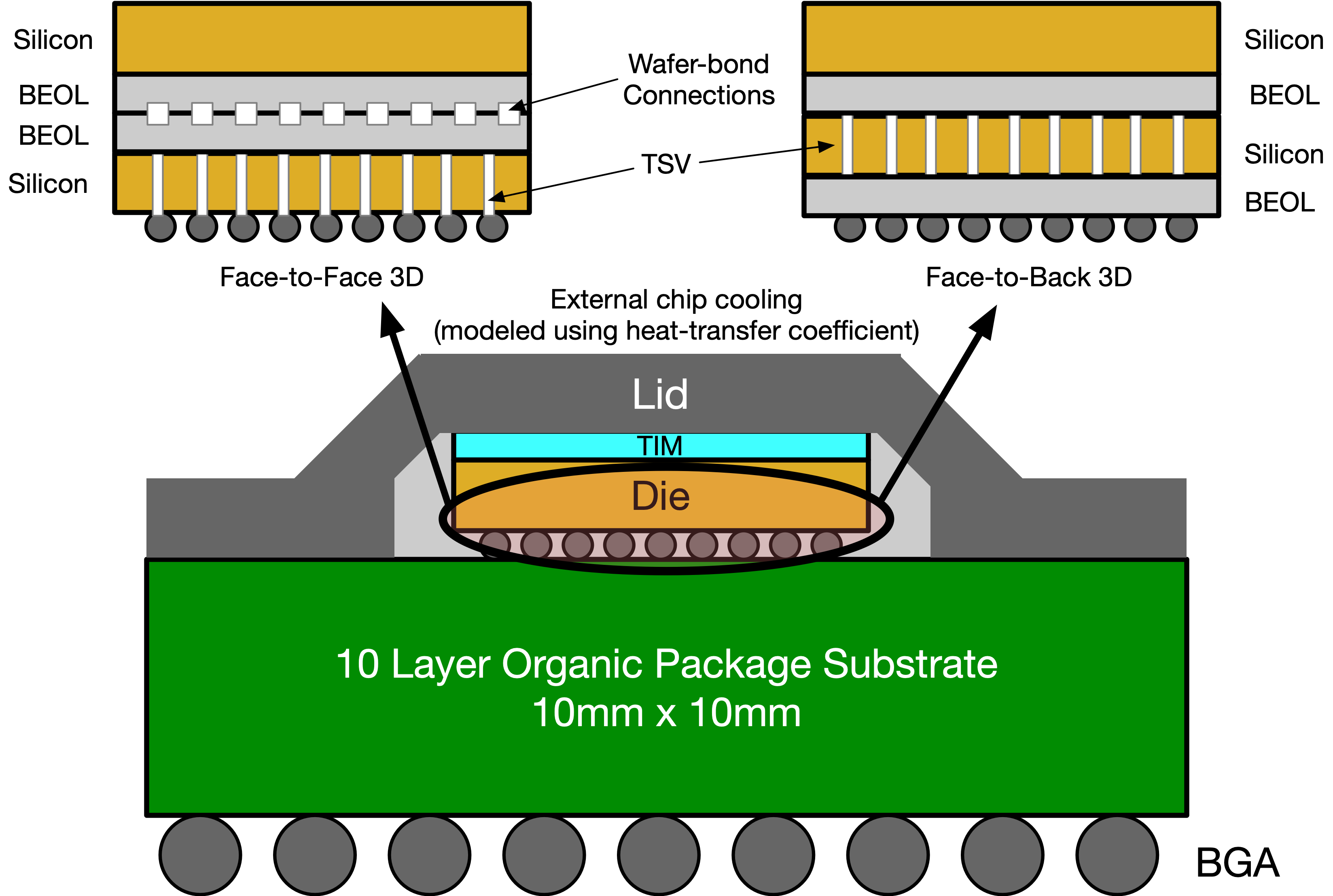}}
\caption{Cross-section of the die and package model used in this study.}
\label{fig_cross_section}
\end{figure}

Figure \ref{fig_cross_section} describes the package and die cross-section model in this study. Face-to-face (F2F) 3D connections are formed using wafer-bond pads at the F2F interface, while TSVs are used to escape the bottom of the die for power delivery and I/O signals. Face-to-back (F2B) 3D connections require TSVs for connection between the two dies. The rest of the paper is organized as follows: Section II explains the implementation flow of 3D microprocessor design. Section III and IV describe the thermal analysis simulation framework and experimental setup. Section V discusses results and provides insights by comparing the thermal characteristics of different 3D configurations. In Section VI, conclusions are drawn for thermal aware design of high-performance microprocessors using 3D integration technologies.

\section{3D CPU implementation}
The 2D reference design is a 64-bit, out-of-order, high-performance Arm\textsuperscript{\textregistered} CPU. The physical design of the CPU was done in 3D, using a novel physical implementation flow which consists of design-aware RTL partitioning and placement co-optimization across multiple tiers. 
%The 3D design can achieve the same frequency as the 2D design but with a larger L2 cache size.

\begin{table}[!t]
\caption{Comparison of 2D and 3D CPU}
\begin{center}
\begin{tabular}{|c|c|c|}
\hline
\textbf{Metric} & \textbf{2D} & \textbf{3D}\\
\hline
Process node & 7nm & 7nm \\
\hline
Maximum frequency & 3 GHz & 3 GHz \\
\hline
L2 capacity (normalized) & 1X & 2X \\
\hline
L2 access latency (normalized) & 1X & 1X \\
\hline
2D Footprint (normalized) & 1X & 0.77X \\
\hline
\hline
\end{tabular}
\label{tab_ares}
\end{center}
\end{table}

\begin{figure}[!b]
\centering
\centerline{\includegraphics[width=3.2in]{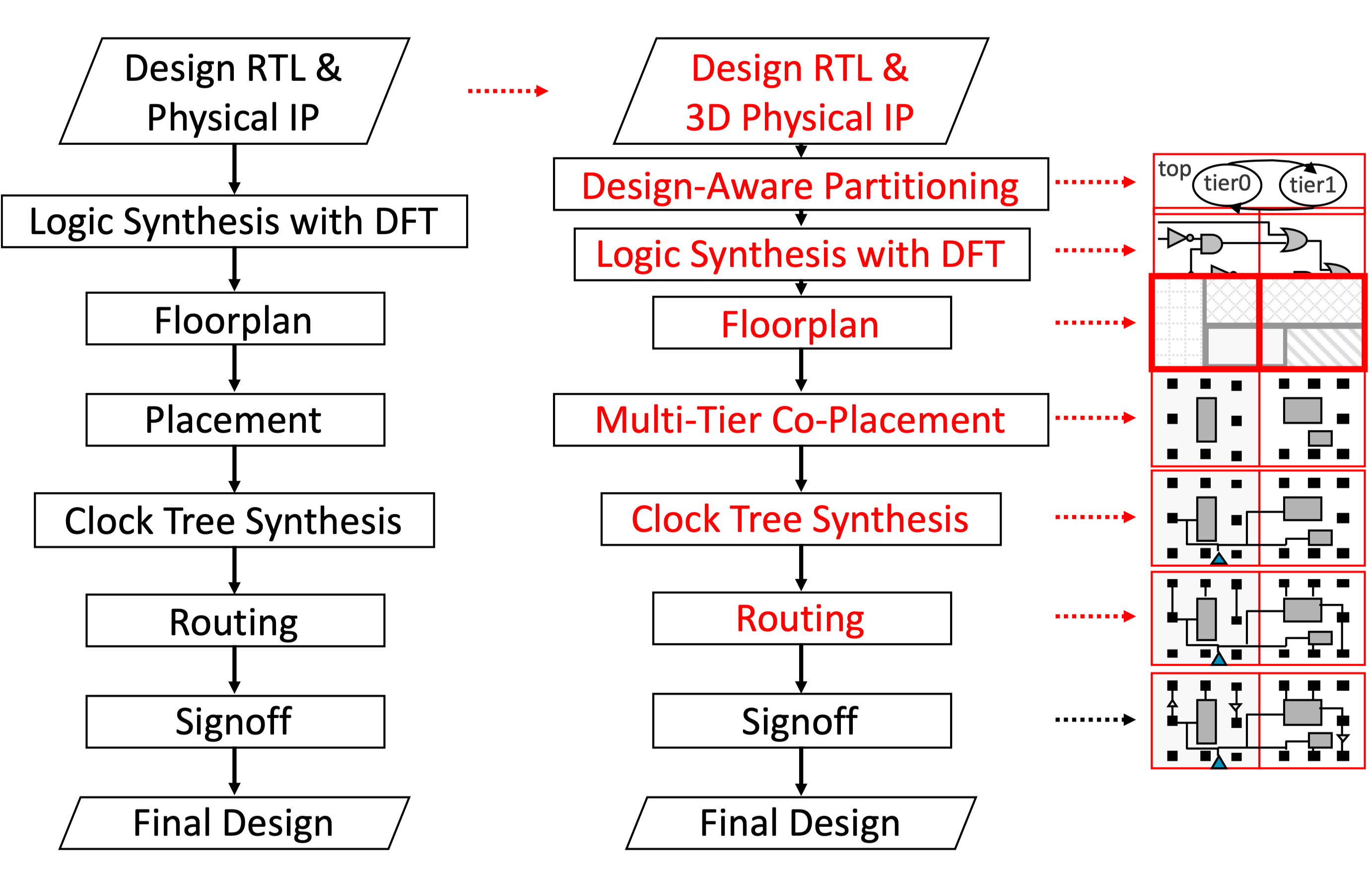}}
\caption{2D (left) and 3D (right) physical implementation flow chart. The steps highlighted in red differ from the standard 2D physical implementation flow.}
\label{fig_3dflow}
\end{figure}

\begin{figure}[!t]
\centering
\centerline{\includegraphics[width=3.45in]{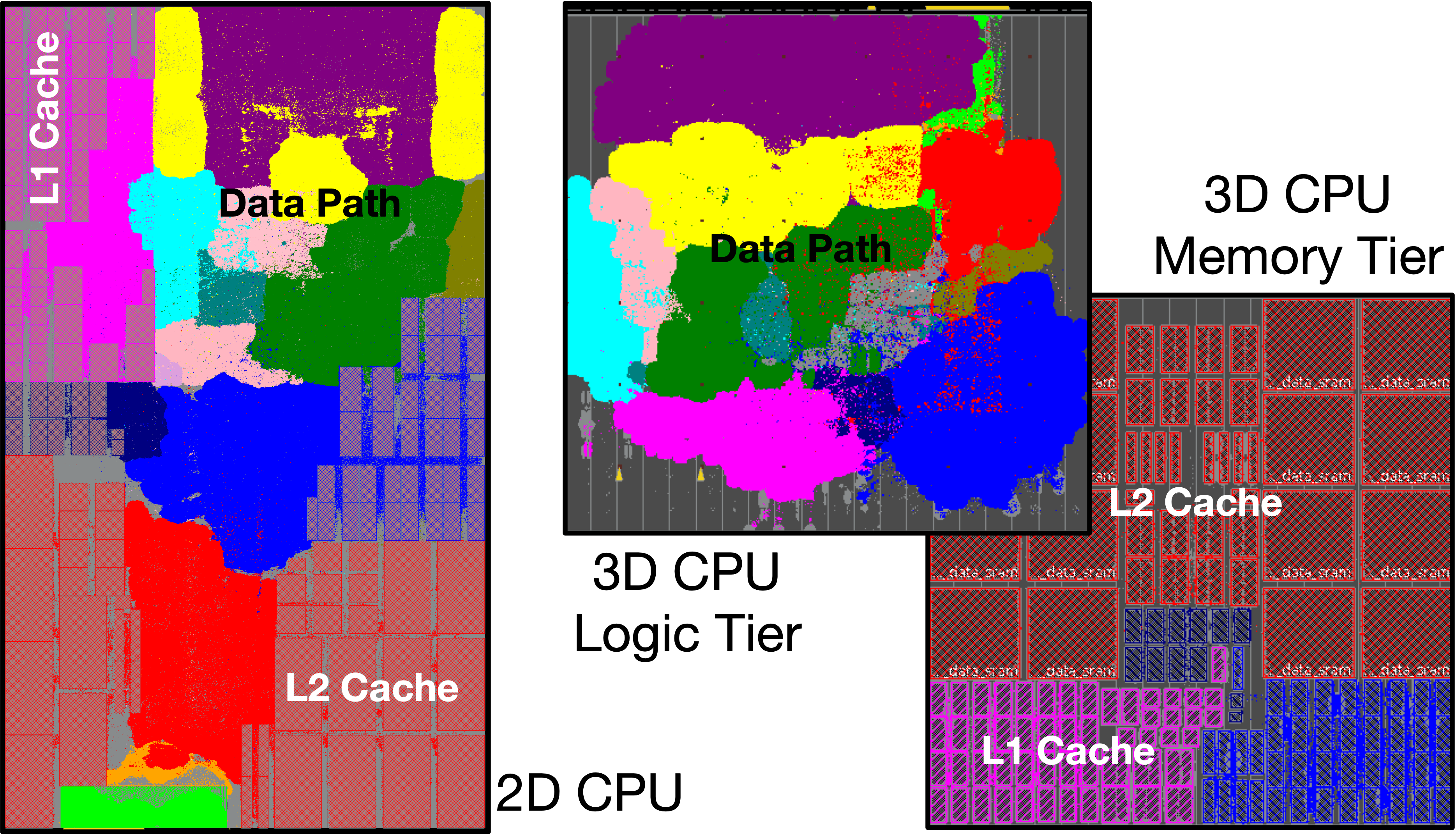}}
\caption{Layout of 2D CPU, logic and memory tiers of the 3D CPU after physical design implementation sign-off.}
\label{fig_impl}
\end{figure}

Specifically, using high-density 3D F2F wafer-bonding, the reference design was partitioned into separate memory and logic tiers. The memory tier consists of Level 1 (L1) and Level 2 (L2) caches, while the logic tier consists of the logic blocks including the CPU datapath. Additionally, the 3D CPU was configured to include twice the L2 cache capacity compared to the 2D design. Increasing the L2 cache capacity in a 2D design results in a larger CPU floor-plan, longer wire-delay and hence, has a higher memory access latency. Since 3D stacking enables bringing compute and memory blocks closer together, the 3D CPU design can accommodate a 2X larger L2 size compared to the 2D design with the same memory access latency. Details of the 2D and 3D CPU designs are provided in Table \ref{tab_ares}.

Utilizing a novel 3D implementation flow, which allows cross-tier placement optimization, the 3D CPU implementation was able to achieve the same target frequency (3 GHz) as the 2D design. The critical steps of the flow are described in Figure \ref{fig_3dflow} and more details are provided in \cite{xu:2019}, \cite{sinha:2019}. This highlights the importance of design-aware partitioning and cross-tier optimization to achieve high-performance 3D design. 
Figure \ref{fig_impl} shows the placed and routed 2D and 3D CPU designs. Since the 3D CPU consists of a 2X larger L2 cache, it has a larger total area compared to the 2D CPU (1.54X total silicon area relative to the 2D design). However, the 2D footprint of the 3D stacked design is approximately 0.77X of the 2D design. 

\section{Thermal analysis framework}

\begin{figure}[!b]
\centering
\centerline{\includegraphics[width=3in]{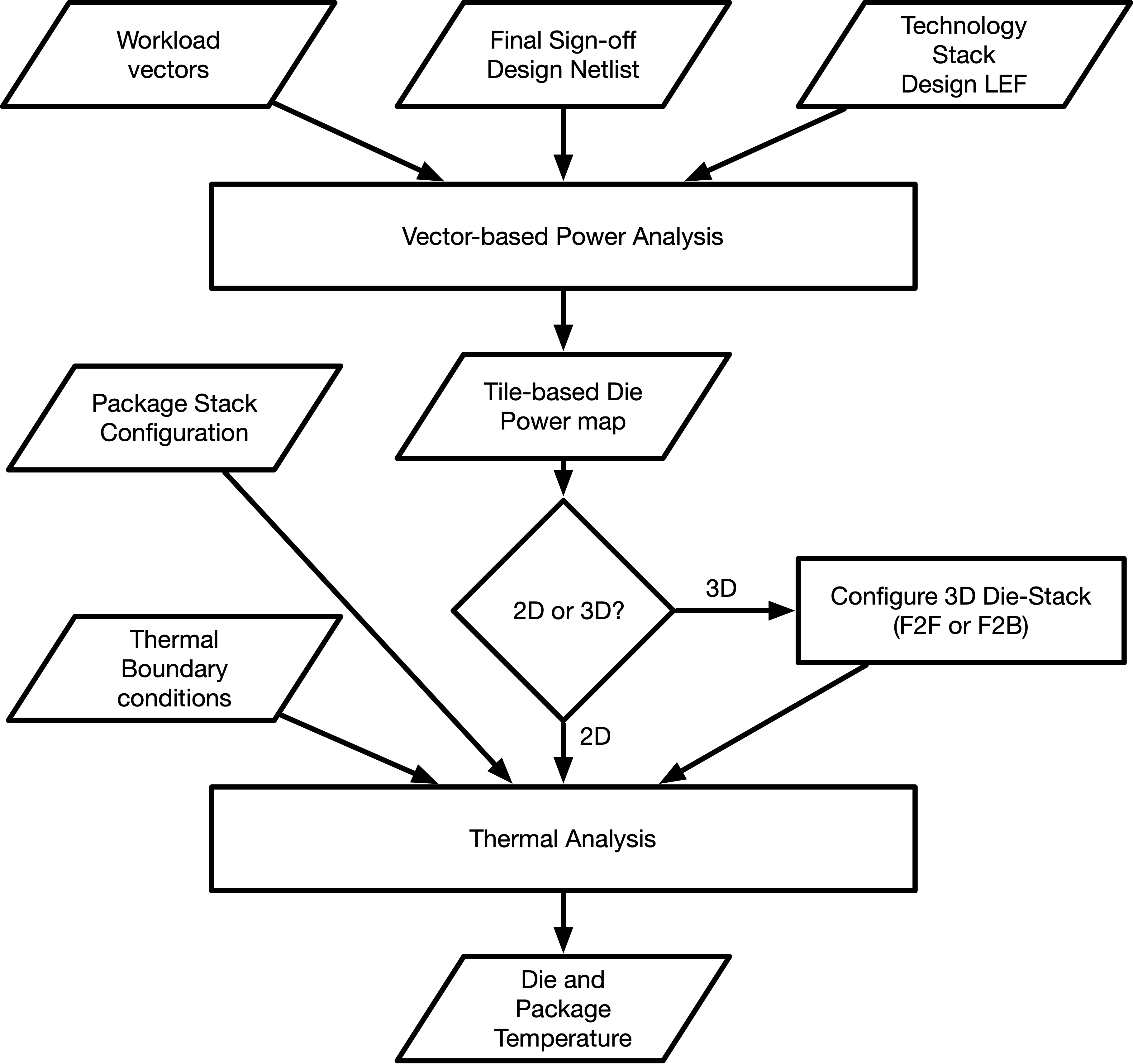}}
\caption{Flowchart for thermal analysis.}
\label{fig_thermal_flow}
\end{figure}

Figure \ref{fig_thermal_flow} outlines the flow used to perform package level thermal simulations on the 2D and 3D implementations of the CPU discussed in Section II. \textit{Cadence\textsuperscript{\textregistered} Celsius{\texttrademark} Thermal Solver} \cite{celsius} was used to perform both static and transient simulations. The tool requires a location-based power dissipation map of the die-model as well as a detailed description of the die stack-up, i.e., devices, interconnects and dielectrics along with their thermal conductivity properties. In this section, we describe the methodology to generate the tile-based power map file for the design for thermal analysis and calibration of the thermal boundary conditions with hardware measurement data.

A Vector-based analysis is done using \textit{Synopsys\textsuperscript{\textregistered} PrimeTime{\texttrademark} PX} \cite{ptpx}, which reports the cell level power consumption of the full design. Two power-indicative workload vectors are utilized to estimate power: 
\begin{itemize}
\item $dhrystone$ is used for characterizing the average power of the CPU. This is indicative of a typical real-world use-case of the CPU. 
\item $maxpower$ is used to characterize the worst-case power dissipation of the CPU. It is important to note that real-world usages typically only have $maxpower$ characteristics for a short duration of time. 
\end{itemize}

\begin{figure}[!t]
\centering
\centerline{\includegraphics[width=3.2in]{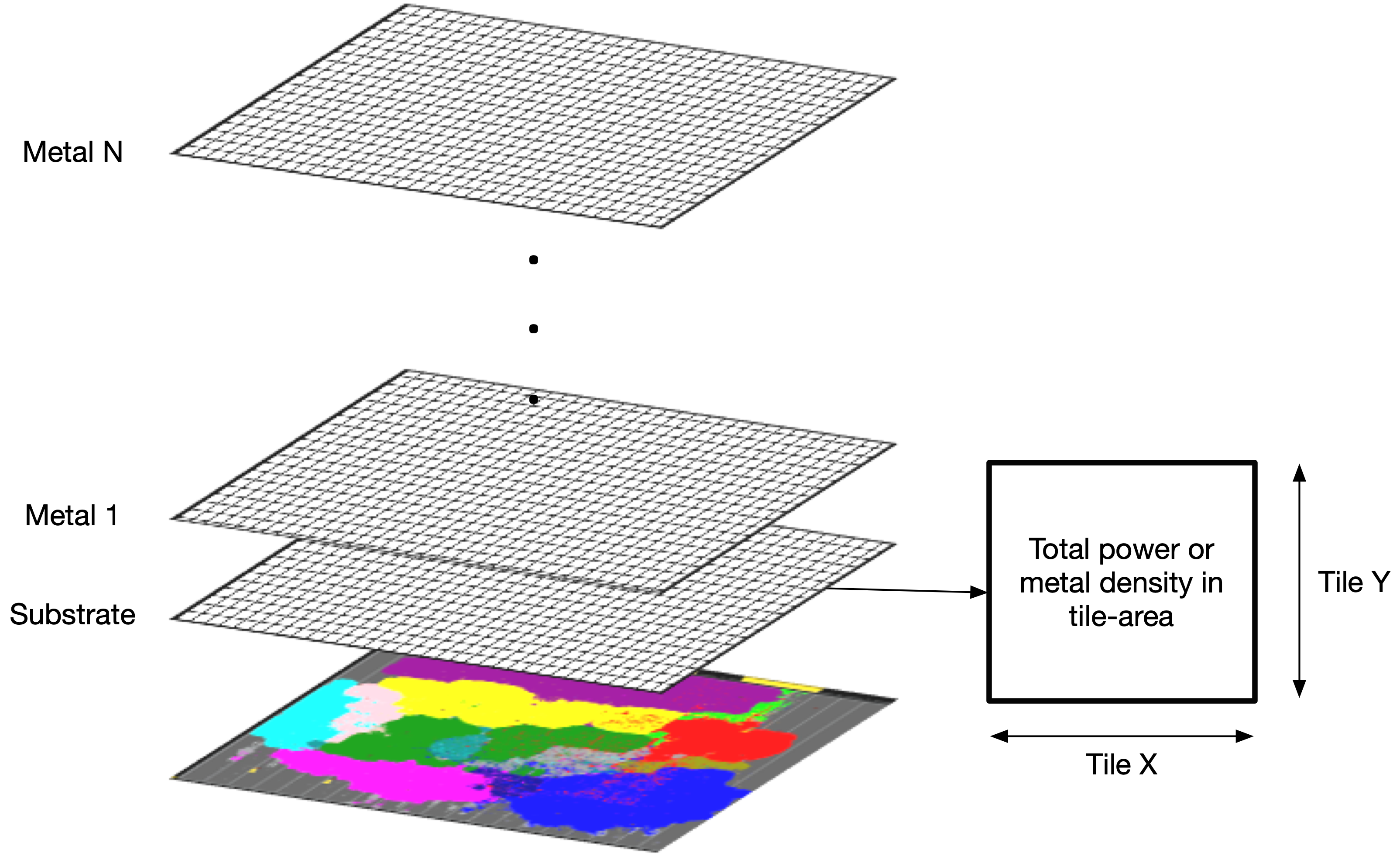}}
\caption{Tile-based power and metal density map used for thermal analysis.}
\label{fig_tile}
\end{figure}

Based on the results of the vector-based power analysis, a tile-based power map of the die is generated. As depicted in Figure \ref{fig_tile}, the entire die is divided into equal-sized tiles. The power of each tile is the sum of the internal power, switching power and leakage power associated with the cells within the tile. The number of tiles for the die was chosen to ensure high resolution of the power density within different modules as well as maintain reasonable computation runtime of the tool. The tile-based power map file also contains metal density and thermal properties of all the layers in the back end of line (BEOL) stack as well as the substrate (Silicon). \textit{Cadence Celsius Thermal Solver} uses the power map file along with a complete description of the package stack-up, bumps, die-stack, molding compound, lid, thermal-interface material (TIM) and boundary conditions. Setting up realistic boundary conditions for thermal analysis is critical for getting accurate results and is described in the next subsection.

\subsection{Model Calibration}
\label{subsec:model_calibration}
\begin{figure}[!t]
\centering
\centerline{\includegraphics[width=3.45in]{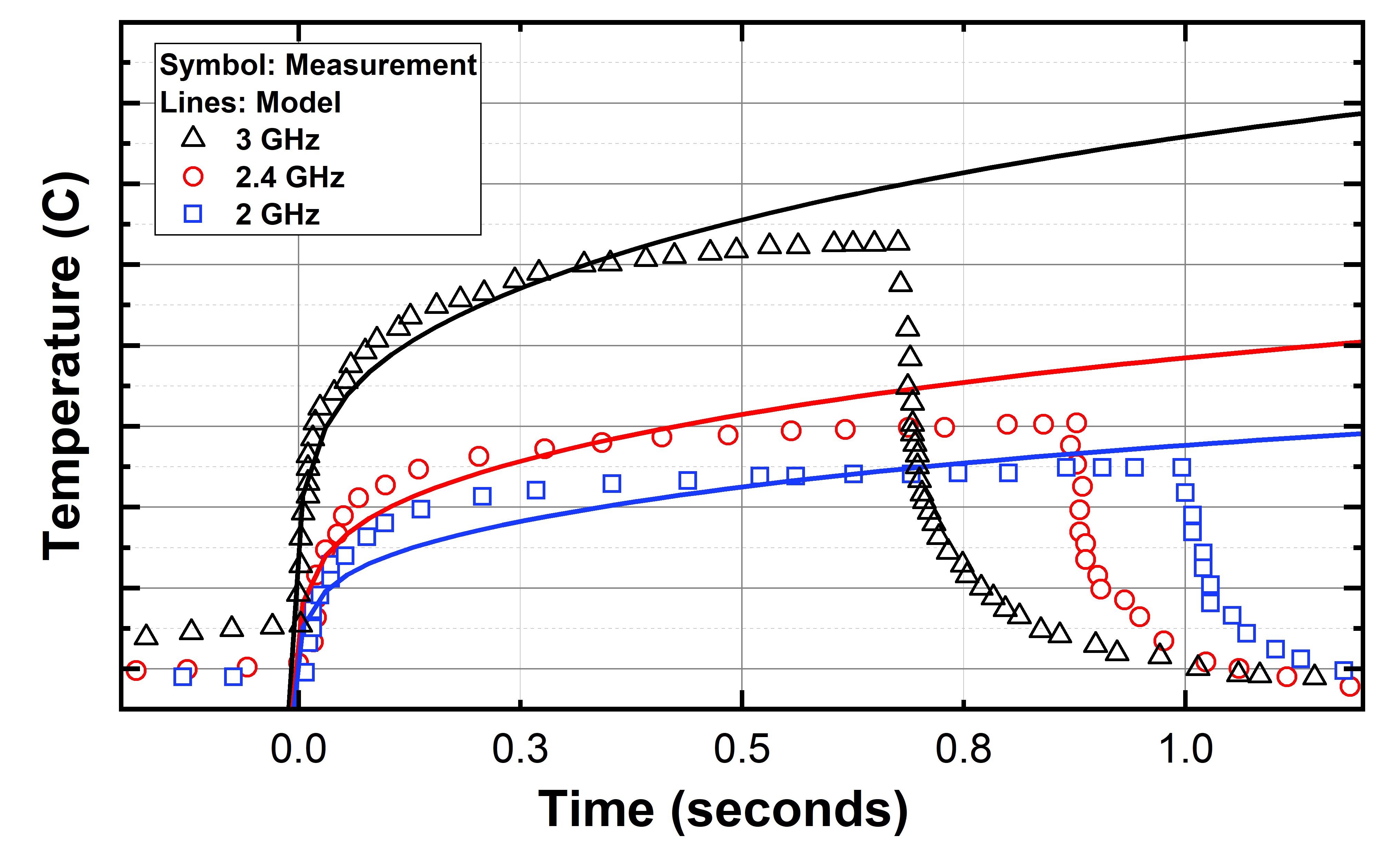}}
\caption{Thermal simulation boundary condition calibration with on-die measurements. The simulation package, die model, workload, and power dissipated were matched with the measurement setup.}
\label{fig_calibration}
\end{figure}

The thermal simulation boundary conditions are calibrated to hardware measurements on a 4-core SoC test chip fabricated on a 7nm process technology \cite{christy:2020}. The $maxpower$ workload was run for a fixed number of CPU cycles on all four cores and temperature measurements from on-die temperature sensors were collected. The power dissipation from the four cores was also measured. The package stack-up and die power map matching the SoC was generated and set up for thermal simulations. The hardware measurement setup consists of a heat sink and a fan on top of the package lid. The heat transfer coefficient (HTC) in transient simulations that provided the best match to the temperature measurement data at the different operating frequencies of the chip was finalized as the boundary condition for all subsequent experiments. Figure \ref{fig_calibration} shows the calibration of our thermal simulation boundary conditions to the temperature sensor measurements.

\section{Experimental setup}

\begin{table}[!b]
\caption{3D Simulation Configurations}
\begin{center}
\begin{tabular}{|c|c|}
\hline
\textbf{Configuration} & \textbf{Description} \\
\hline
\hline
\# of CPU in cluster & 2 \\
F2F & Face-to-Face 3D \\
F2B & Face-to-Back 3D \\
Logic-on-Mem & Logic on top and memory on bottom tier \\
Mem-on-Logic & Memory on top and logic on bottom tier \\
CPU-on-CPU & 2D CPU on top and bottom tier \\
Spacing between CPUs & 200 $\mu$m \\
Margin around CPU cluster & 1 mm\\ 
Package dimensions & 10mm x 10mm \\
\hline
\hline
\end{tabular}
\label{tab_3dconfig}
\end{center}
\end{table}

\begin{figure}[!t]
\centering
\centerline{\includegraphics[width=3.45in]{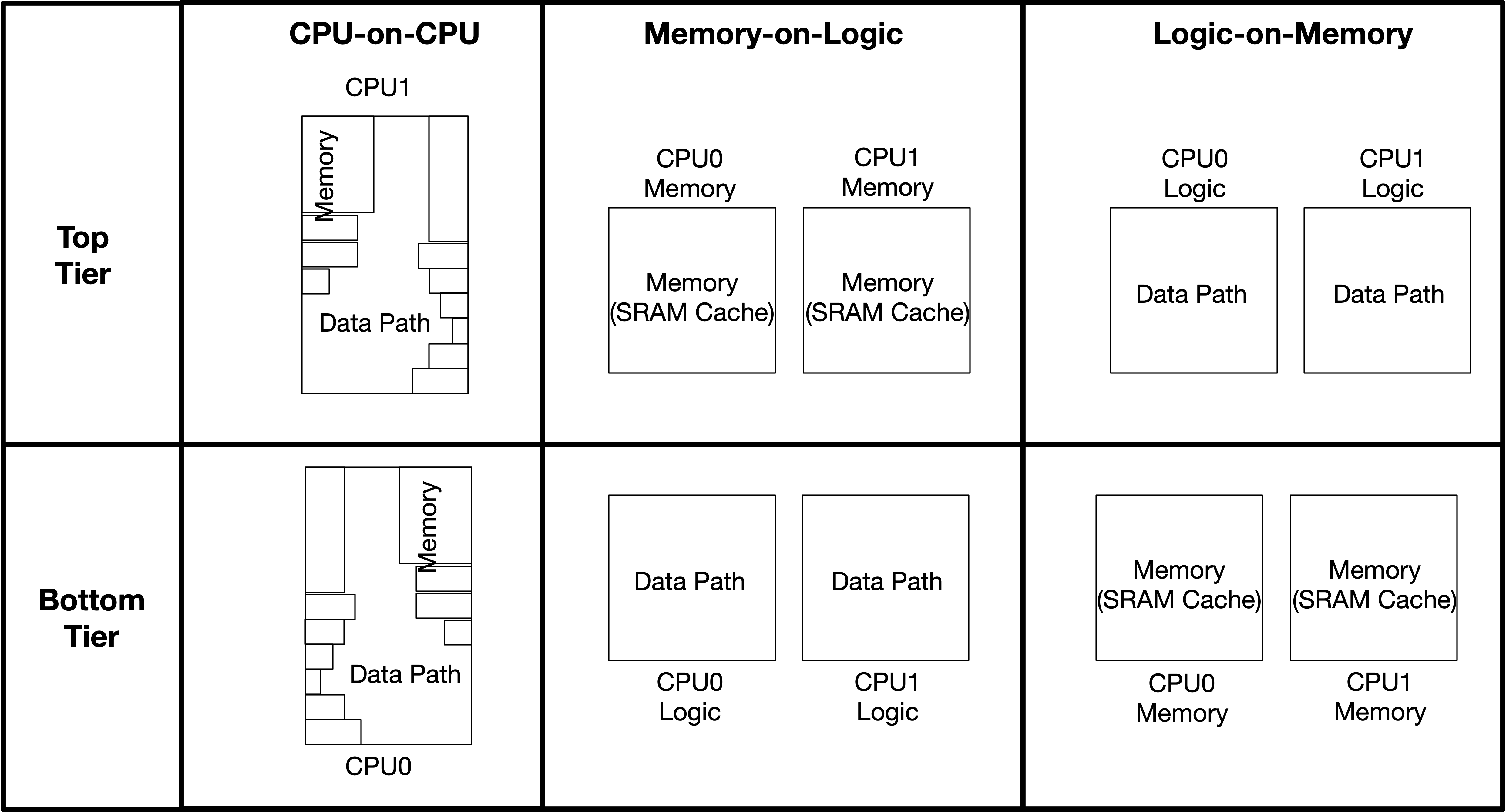}}
\caption{Pictorial view of the top and bottom tiers of the 3D simulation configurations.}
\label{fig_config}
\end{figure}

A dual-core CPU cluster was used for our thermal simulation study with the same workload running on both CPUs. It is important to note that the physical implementation of the 3D CPU requires high-density 3D connections in a F2F configuration. However, abstracting the CPU power in terms of tile-based power maps allows us to explore other different 3D configurations such as F2B 3D as well. TSVs are modeled at appropriate locations in the die stack-up, for example, between the bottom-die and the package for F2F configuration; and between bottom-die and top-die in F2B configuration (see Figure \ref{fig_cross_section}). For each 3D configuration, logic and memory dies were swapped to study the impact of the proximity of die to external cooling on top of the package. Additionally, a CPU-on-CPU 3D stack is modeled, where each die consisted of a single-core 2D CPU. This represents the current state-of-the-art where functionally complete `chiplets' are stacked in 3D. The analysis of all these configurations is done under the similar die and package size assumptions. The two CPUs are placed 200 $\mu$m apart with a margin of 1 mm to the die edges. The static power from a representative system-level cache is allocated to the margin area to model idle caches placed around the CPUs. A 10x10 mm$^2$ package comprising of 10 build-up layers was used for all configurations. All temperature values are reported relative to the dual-core 2D die. Table \ref{tab_3dconfig} lists the nomenclature for the different 3D configurations used in our experiments and Figure \ref{fig_config} provides a pictorial representation of how they are organized on the two tiers.

\begin{figure*}[!t]
\centering
\includegraphics[width=\textwidth]{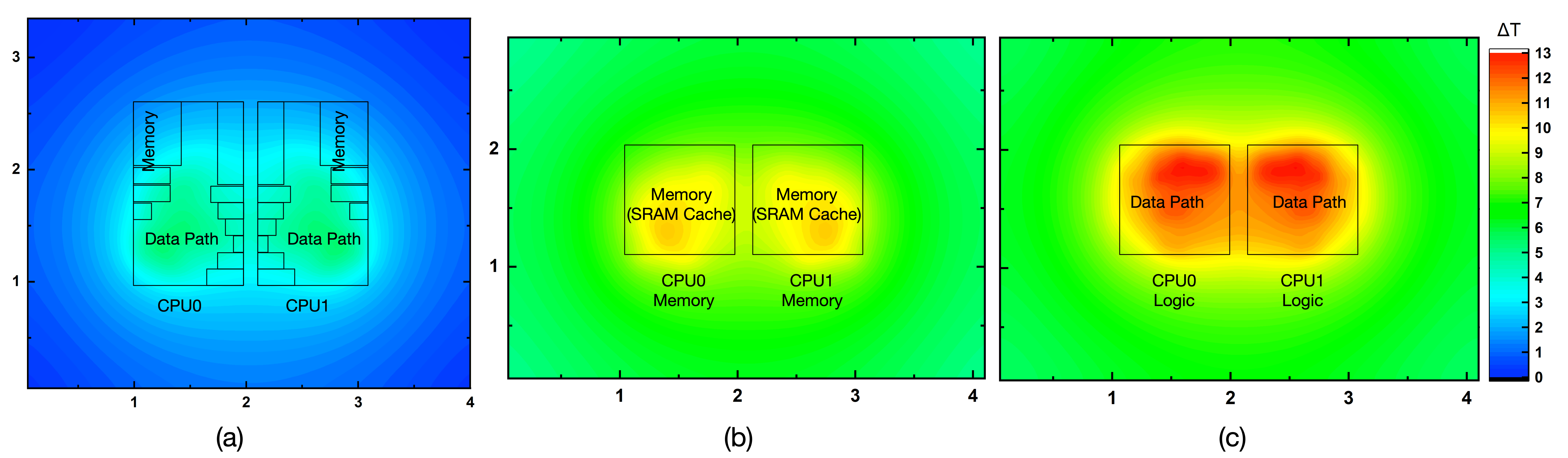}
\caption{Heat-maps under the $maxpower$ workload for (a) a 2D dual CPU configuration, (b) a 3D dual CPU memory tier and (c) a 3D dual CPU logic tier. The die dimensions are in mm. The temperature values are relative to the coolest point on the 2D die. In this scenario, the 3D stack has the logic tier on top (close to the heat sink) and the memory tier at the bottom (close to the package). }
\label{fig_heatmaps}
\end{figure*}

\section{Results}
The steady-state temperature heat map of a baseline 2D dual-core CPU configuration and the two tiers of the 3D CPU in F2F Logic-on-Memory configuration are shown in Figure \ref{fig_heatmaps}. The temperature values are relative to the coolest point on the 2D die. Each CPU core is running the $maxpower$ workload. The heat map clearly emphasizes that in the 2D CPU, the data-path runs hotter ($\Delta T \approx 6\degree$C) than the L1 and L2 cache blocks. This observation directly correlates to the 3D logic-over-memory heat maps, where the logic die is hotter than the memory die despite being in closer proximity to the heat sink.

\begin{figure}[!b]
\centering
\centerline{\includegraphics[width=3.45in]{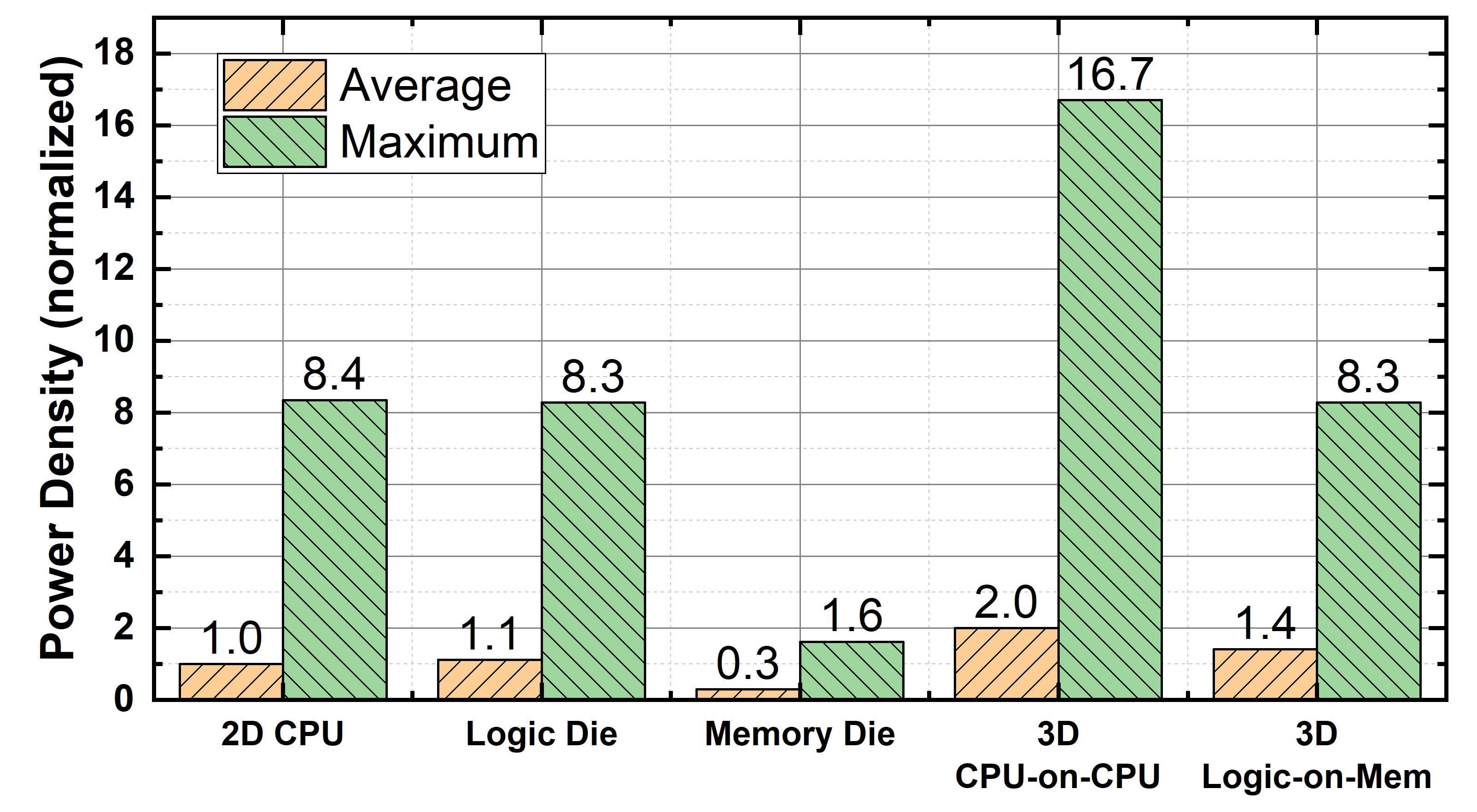}}
\caption{Maximum and average power density in the 2D and 3D dies for different configurations running the $maxpower$ vector.}
\label{fig_power_density}
\end{figure}

To gain further insights into the temperature profiles, the maximum and average power density of each die in the 2D and 3D configurations are plotted in Figure \ref{fig_power_density}. Even though the maximum power density on the die is similar between the 2D and the 3D logic-over-memory case, the average power density of the 3D stack is higher, owing to a similar total power of the two designs in a 0.77X smaller footprint. This results in a higher temperatures in the 3D stacked designs.

\begin{figure}[!t]
\centering
\centerline{\includegraphics[width=3.45in]{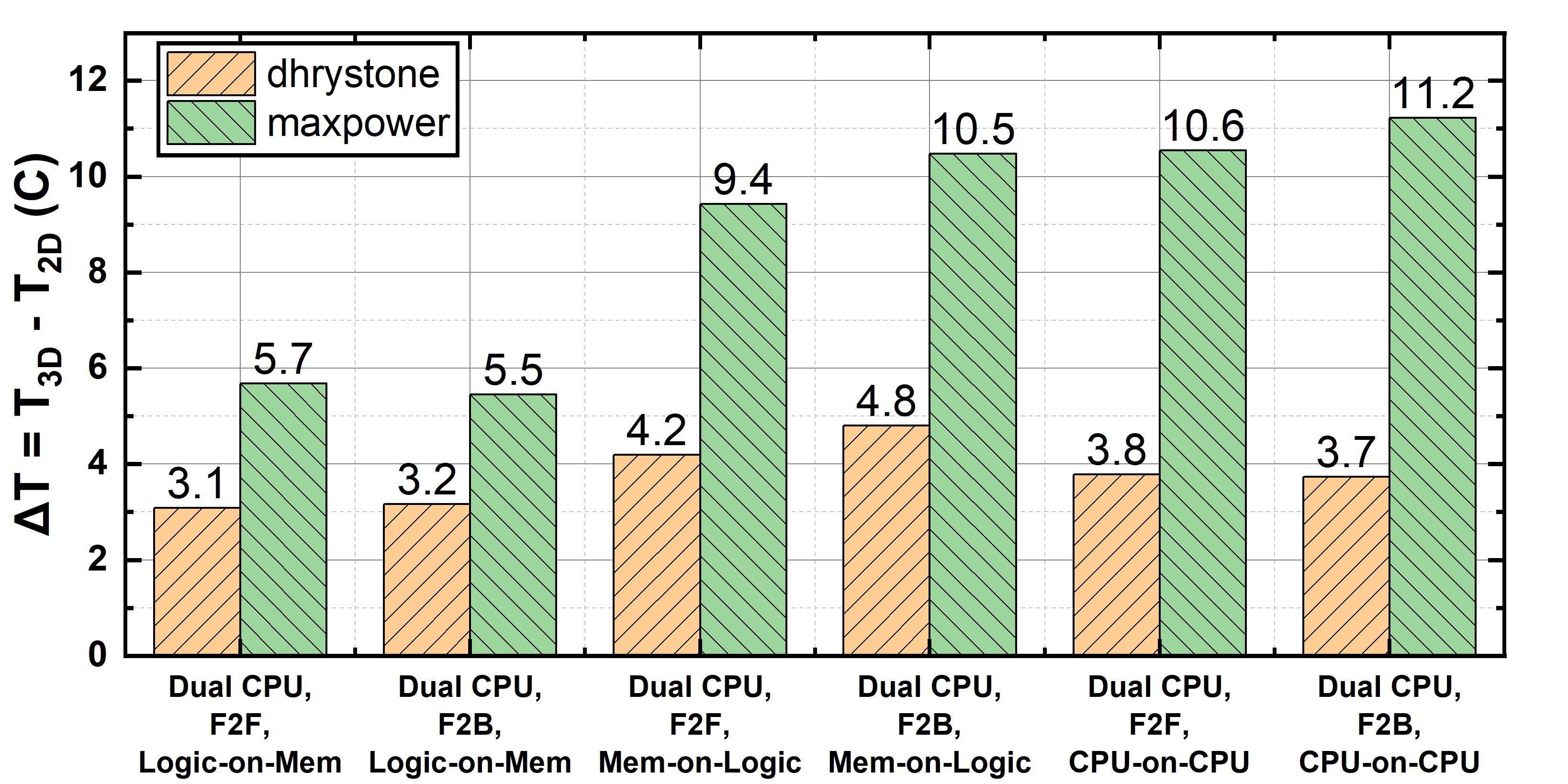}}
\caption{Increase in maximum temperature for different 3D configurations under the $dhrystone$ and $maxpower$ workloads relative to 2D baseline.}
\label{fig_bar_graph}
\end{figure}

Figure \ref{fig_bar_graph} plots the maximum steady-state temperature (relative to the 2D baseline) for all the 3D configurations running $dhrystone$ and $maxpower$ workloads. For the 3D logic and memory partitioned CPU, the logic tier is always hotter due to higher power density. For the CPU-on-CPU case, the CPU on the bottom die shows a higher temperature profile since it is not in proximity to the heat sink.

Among the different 3D configurations, the logic-on-memory 3D design has a temperature rise of around 5$\degree$C while the memory-on-logic 3D design has a temperature rise of 9$\degree$C compared to the 2D baseline. This is because the higher power logic die in the memory-on-logic configuration sits away from the heat sink. CPU-on-CPU has the worst thermal characteristics because of overlapping hotspots from the two tiers, where the maximum and average power density doubles in the CPU-on-CPU configuration compared to a 2D CPU as shown in Figure \ref{fig_power_density}. 

\begin{figure}[!b]
\centering
\centerline{\includegraphics[width=3.2in]{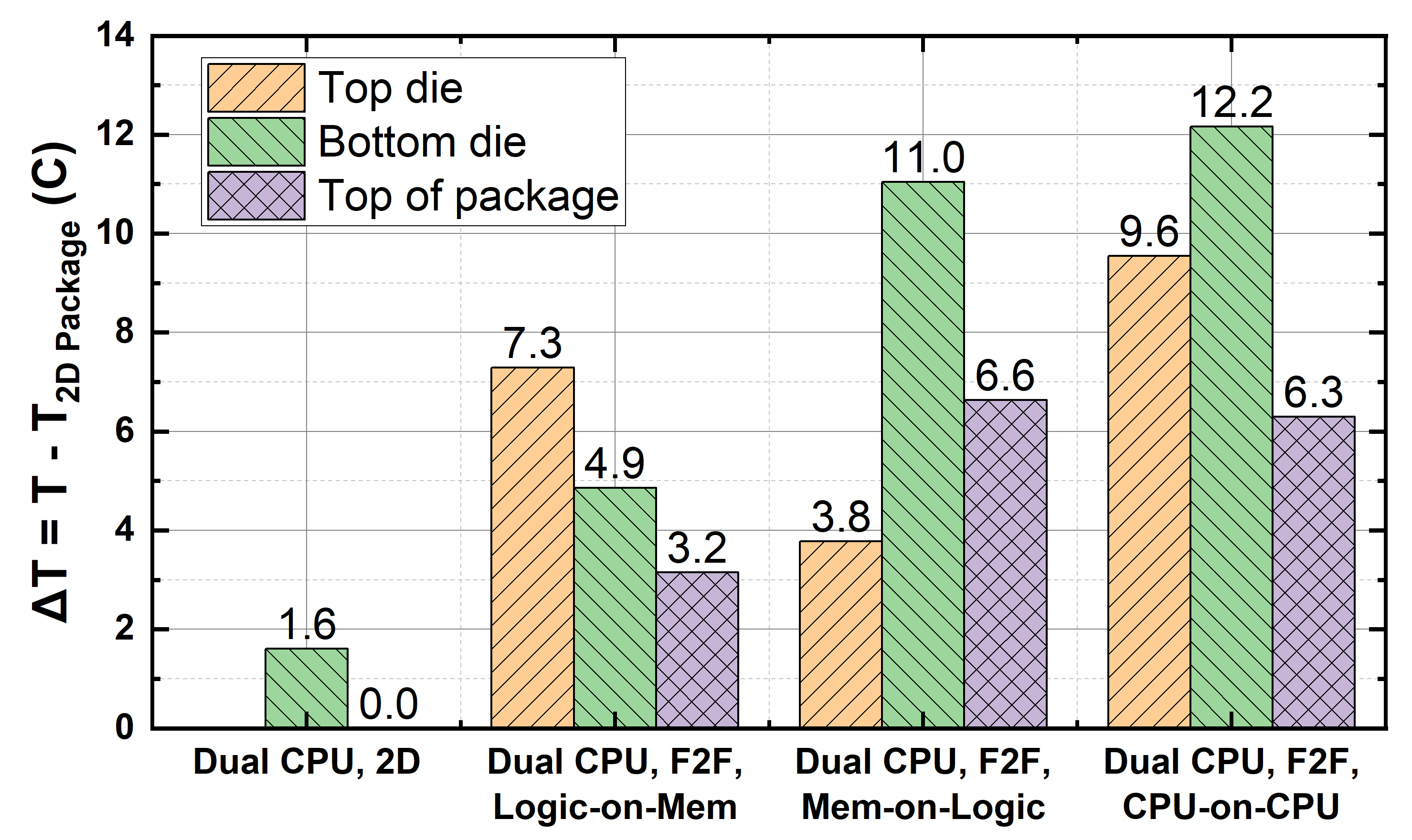}}
\caption{Package and die temperatures for different 3D configurations under the $maxpower$ workload compared to 2D package temperature as baseline.}
\label{fig_bar_graph_pkg}
\end{figure}

F2F and F2B 3D configurations have a minor impact on the temperature profile, primarily because in each case the bottom die is thinned down to process TSVs, providing a lower effective thermal resistance to the package. To the first order, the temperature rise in 3D stacking is mainly proportional to the effective power density of the design. Figure \ref{fig_bar_graph_pkg} shows the maximum temperatures on both dies as well as the package across different 3D configurations under the $maxpower$ workload relative to the 2D package temperature, which highlights the value of logic-over-memory 3D design in minimizing the temperature rise in the package. 

\begin{figure}[!t]
\centering
\centerline{\includegraphics[width=3.2in]{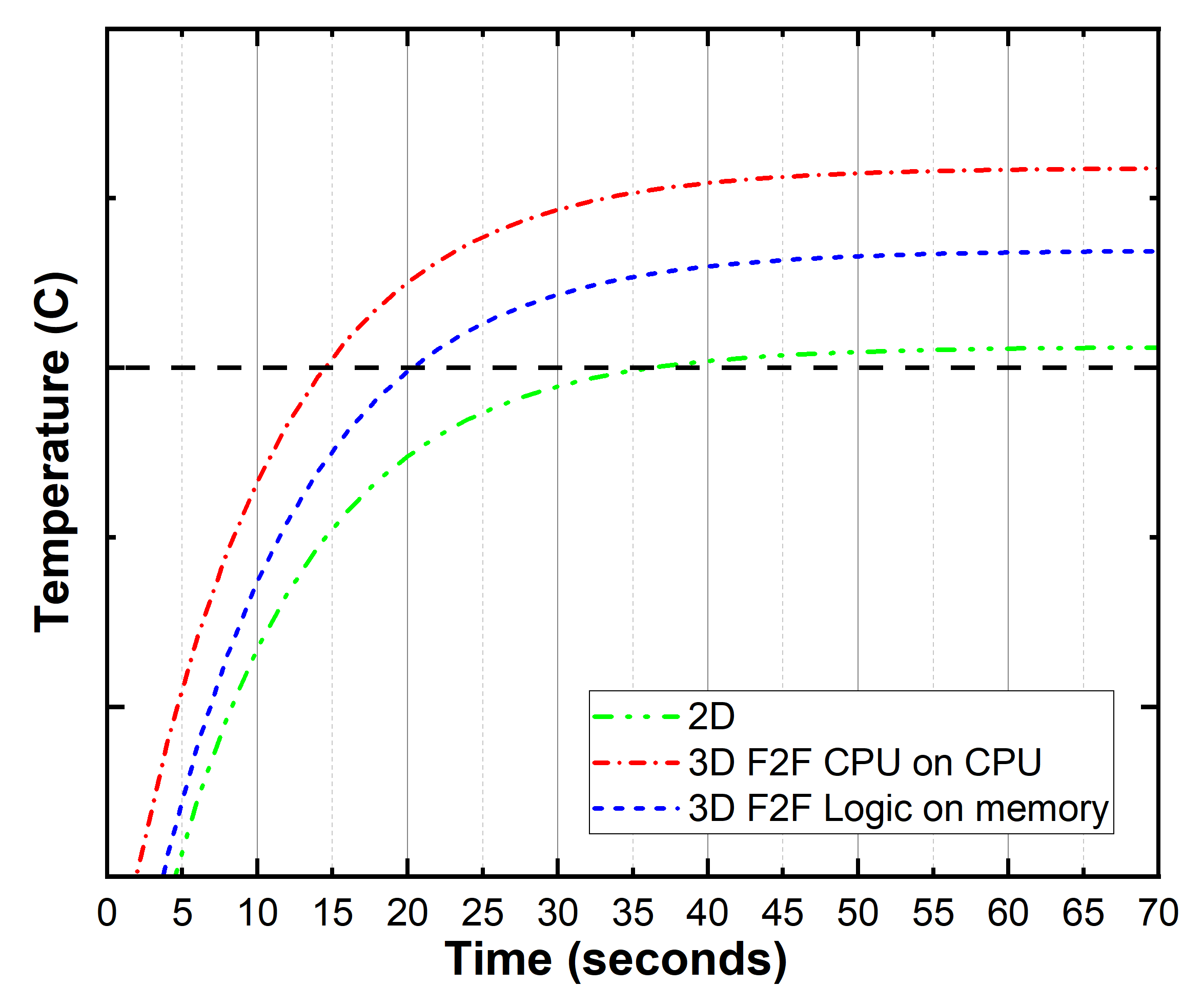}}
\caption{Transient profile of 2D and 3D configurations under the $maxpower$ workload.}
\label{fig_transient}
\end{figure}

Self-heating of CPUs is a relatively slow process (seconds) compared to the operating frequency of the design (sub-nanoseconds). Hence, the steady-state temperature profiles provide a limited view of the design or performance trade-offs of 2D and 3D stacked systems. Depending on the time required to reach maximum allowed operating temperature, the power profile of the application and the application runtime, the performance of 3D systems may need to be throttled to stay within the thermal constraints. Figure \ref{fig_transient} shows the transient temperature rise for the 2D and 3D stacked configurations under the $maxpower$ workload. Assuming the dashed horizontal line corresponds to the maximum allowed temperature for the chip, the 2D CPU can run for 35 seconds before performance needs to be throttled to allow the system to cool down. However, the 3D logic-on-memory design and 3D CPU-on-CPU design can run only for 20 and 15 seconds respectively (a 0.57X and 0.43X reduction). It is important to highlight that this describes a worst-case scenario for power consumption. We expect that more mainstream workloads featuring average power dissipation can run sustainably with appropriate cooling techniques in 3D. We quantify two known approaches to reduce maximum on-die temperature using the worst-case 3D stacking option, CPU-on-CPU configuration.

\begin{figure}[!t]
\centering
\centerline{\includegraphics[width=3.2in]{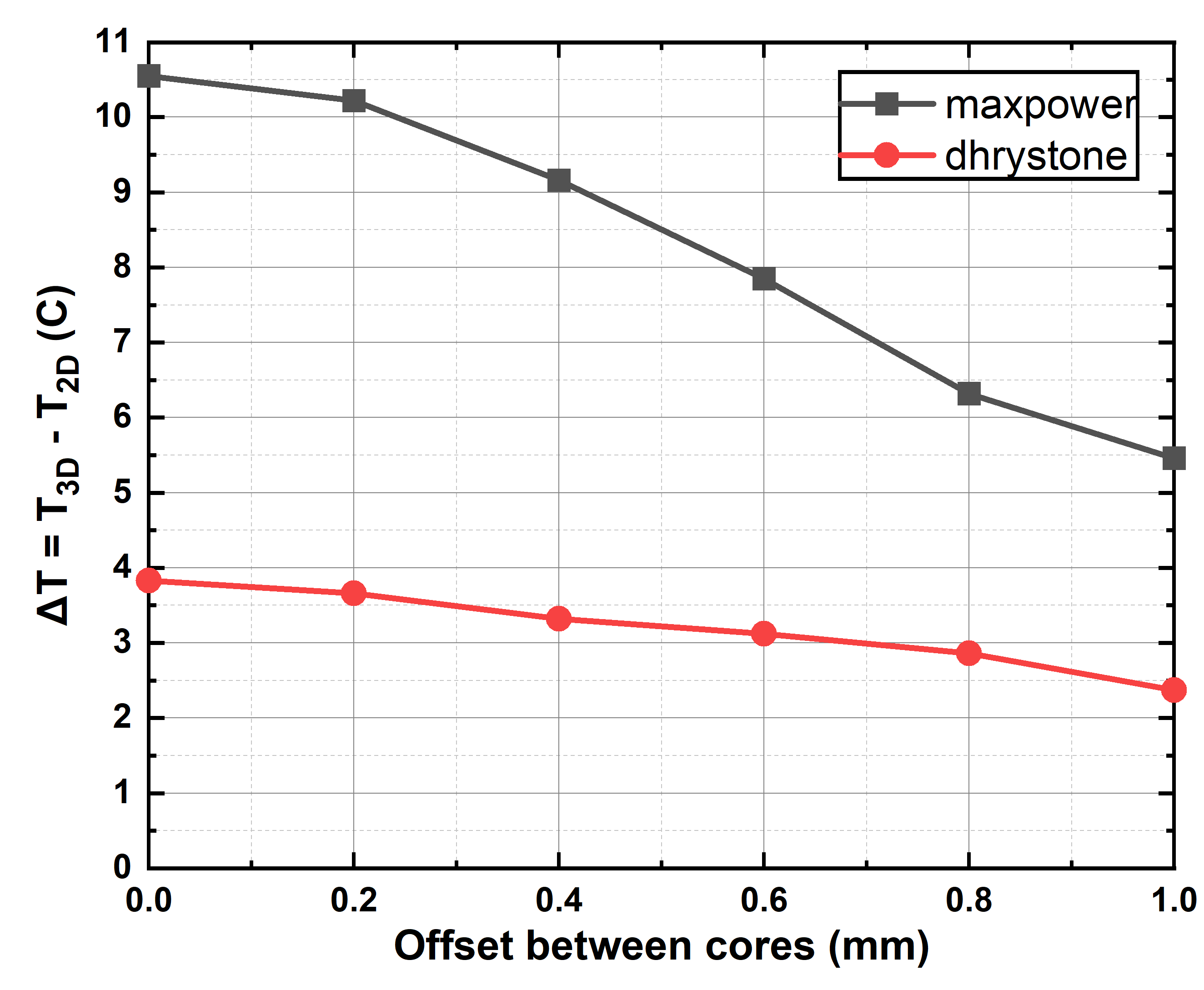}}
\caption{Effect of offset between two tiers on maximum temperature for dual CPU, F2F, CPU-on-CPU under the $dhrystone$ and $maxpower$ workloads compared to 2D baseline.}
\label{fig_offset}
\end{figure}

The CPU-on-CPU 3D configuration is closer to the current state-of-the-art stacking of functionally complete `chiplets'. It does not require design-aware partitioning and timing co-optimization across 3D tiers. However, as seen in the results, this approach has the worst thermal characteristics because the power density of the 3D stack doubles. An approach to mitigate this effect can be to introduce a physical offset between the CPUs on the two tiers and remove overlapping thermal hotspots. Figure \ref{fig_offset} shows that the maximum temperature of this configuration can be reduced by around 5$\degree$C for $maxpower$ and 2$\degree$C for $dhrystone$ when the offset is swept from 0 to 1 mm.

It should be noted that while the method of introducing an offset works for a CPU-on-CPU configuration, it is not feasible for the 3D configurations with logic and memory partitioning since the design utilizes the 3D proximity to reduce latency, enable a larger cache capacity, and close timing at 3 GHz.

\begin{figure}[!t]
\centering
\centerline{\includegraphics[width=3.2in]{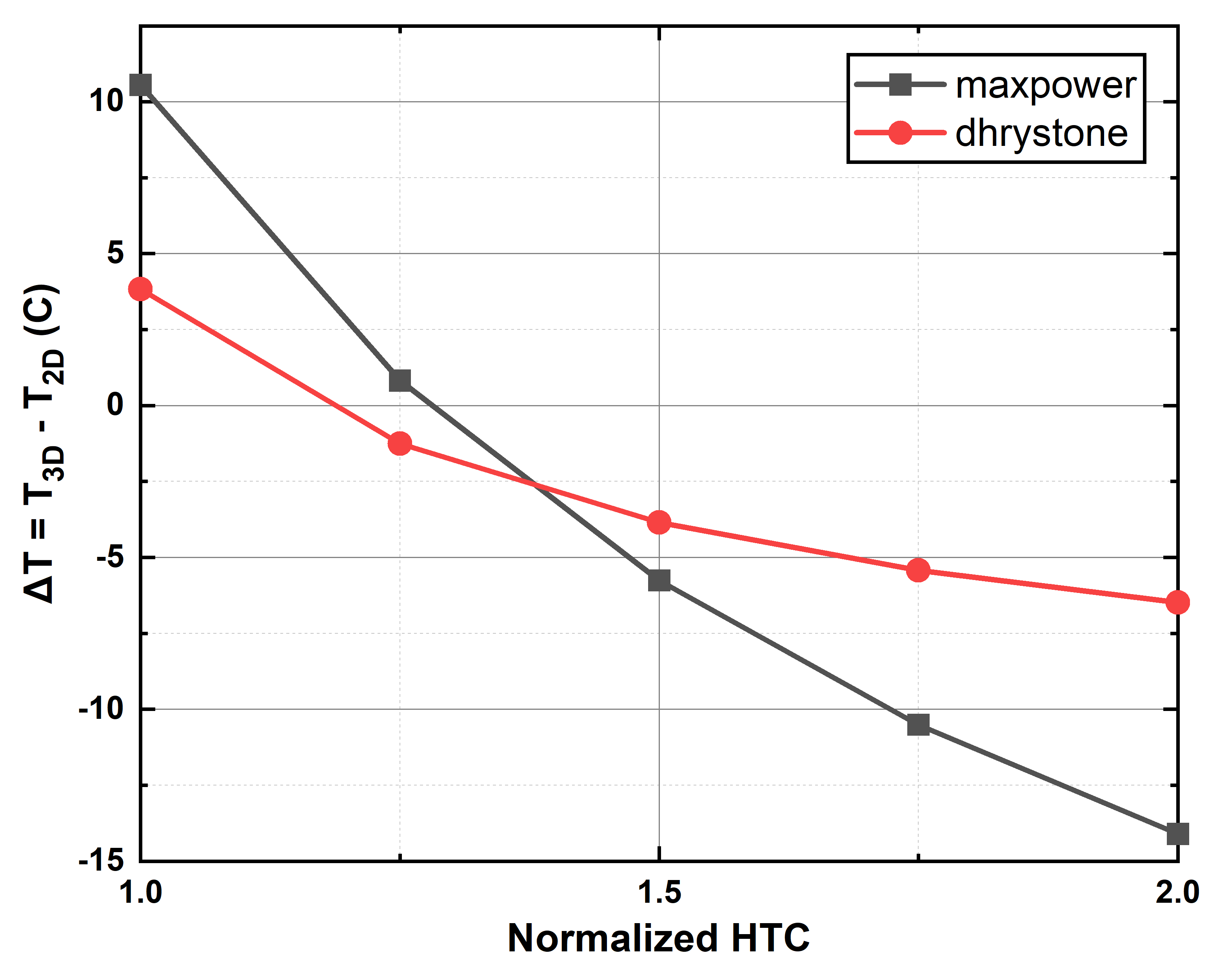}}
\caption{Effect of increasing HTC on maximum temperature for dual CPU, F2F, CPU-on-CPU under the $dhrystone$ and $maxpower$ workloads compared to 2D baseline.}
\label{fig_htc}
\end{figure}

Another approach for reducing die temperatures for 3D could be to use more sophisticated cooling techniques, such as liquid cooling \cite{lau20153d}. Figure \ref{fig_htc} shows the impact of increasing the heat transfer coefficient (HTC) from the hardware calibrated data, on the maximum temperature of a F2F 3D stack CPU-on-CPU configuration. Increasing the HTC on the top of the package i.e., improved cooling is a more effective way of controlling the temperature rise with 3D. It is worth clarifying that the absolute decrease in temperature is more pronounced for the $maxpower$ workload compared to $dhrystone$ because the steady-state maximum temperature from the ambient for the $maxpower$ workload is much higher than the $dhrystone$ workload. 

It is expected that 3D stacked designs will have a relatively lower thermal headroom compared to 2D designs due to higher power density. Through our analysis, it is clear that thermal-aware design methodologies, especially tightly integrated with the physical design flow, will have to be deployed and carefully used in order to get 3D systems to stay within thermal operating budgets, or else in-situ control techniques like DVFS and throttling may have to be employed more at runtime, compared to 2D. It is shown that logic-over-memory partitioned 3D designs provide the best trade-off to manage the aggravated thermal impact. Using improved cooling techniques can further mitigate this issue and thermal challenges are not a showstopper for the adoption of 3D integration technologies. 

\section{Conclusion}
In this work, a comprehensive thermal simulation study of a 3D stacked high-performance CPU using F2F wafer bonding technology was presented. The physical design of the 3D CPU was partitioned into logic and memory tiers and was implemented using a novel physical implementation flow that allows cross-tier timing and placement optimization. Workload-based power and thermal simulation analysis of multiple 3D configurations was performed and compared. All 3D stacked designs show higher maximum temperature due to higher power density. However, it is found that the logic-over-memory 3D stacked configuration is the least impacted, thermally, resulting in a maximum temperature rise of only 6$\degree$C for a worst-case power workload, compared to the 2D baseline. Hence, if thermal-aware design techniques are employed, we emphasize that a relatively small increase in temperature is not a showstopper for 3D integration technologies. This work paves the pathway for future thermal-aware 3D designs.

\section*{Acknowledgment}
The authors will like to thank Heath Perry and Alex Wasson from Arm for providing CPU temperature measurement data and also thank Robert Christy, Rob Aitken and Brian Cline from Arm for technical feedback on the manuscript.

\bibliographystyle{unsrt2authabbrvpp}
\bibliography{ref1}

\end{document}